\documentclass[aps,twocolumn,superscriptaddress,groupedaddress]{revtex4}  % for review and submission

\usepackage{mathrsfs,amsmath,amssymb,amsthm}
\usepackage{amssymb,fmtcount}
\usepackage{graphicx,epsfig,latexsym,overpic,amssymb,color}
\begin{document}
\title{
Bayesian Data Fusion of Imperfect Fission Yields for Augmented Evaluations
}
\author{Z.A. Wang}
\affiliation{
State Key Laboratory of Nuclear Physics and Technology, School of Physics,
Peking University, Beijing 100871, China
}
\author{J.C. Pei}\email{peij@pku.edu.cn}
\affiliation{
State Key Laboratory of Nuclear Physics and Technology, School of Physics,
Peking University, Beijing 100871, China
}
\author{Y.J. Chen}
\affiliation{
China Nuclear Data Center, China Institute of Atomic Energy, Beijing 102413, China
}
\author{C.Y. Qiao}
\affiliation{
State Key Laboratory of Nuclear Physics and Technology, School of Physics,
Peking University, Beijing 100871, China
}
\author{F.R. Xu}
\affiliation{
State Key Laboratory of Nuclear Physics and Technology, School of Physics,
Peking University, Beijing 100871, China
}
\author{Z.G. Ge}
\affiliation{
China Nuclear Data Center, China Institute of Atomic Energy, Beijing 102413, China
}
\author{N.C. Shu}
\affiliation{
China Nuclear Data Center, China Institute of Atomic Energy, Beijing 102413, China
}

%\author{Min-Liang Liu}
%\affiliation{
%Institute of Modern Physics, Chinese Academy of Sciences, Lanzhou 730000,  China
%}
%\author{Fu-Rong Xu}
%\affiliation{
%School of Physics, Peking University, Beijing 100871,                      China
%}
%\affiliation{
%Institute of Theoretical Physics, Chinese Academy of Sciences, Beijing   100080,
%                                                                           China
%}
%\affiliation{Center of Theoretical Nuclear Physics, National Laboratory of Heavy
%                                           Ion Collisions, Lanzhou 730000, China
%}
%123456789 123456789 123456789 123456789 123456789 123456789 123456789 123456789
\begin{abstract}

We demonstrate that Bayesian machine learning can be used to treat the vast amount of experimental
fission data which are noisy, incomplete, discrepant, and correlated.
As an example, the two-dimensional cumulative fission yields (CFY) of neutron-induced fission of $^{238}$U are evaluated with
energy dependencies and uncertainty qualifications.
For independent fission yields (IFY) with very few experimental data, the heterogeneous data fusion
 of CFY and IFY is employed to interpolate the energy dependence.
This work shows that Bayesian data fusion can facilitate the further utilization of imperfect raw nuclear data.

%1234567890123456789012345678901234567890123456789012345678901234567890123456789
\end{abstract}
%\pacs{  21.10.Re, 21.60.Cs, 21.60.Ev}
\maketitle
%123456789 123456789 123456789 123456789 123456789 123456789 123456789 123456789

%\emph{Introduction.}---
%%\label{sec:level1}
%123456789 123456789 123456789 123456789 123456789 123456789 123456789 123456789

Accurate modeling of nuclear fission remains an open problem although its significant applications are generally known~\cite{future}.
In fact,  for modern nuclear energy productions in a more compact and sustainable, cleaner and safer way, a supply of high quality data of multiple fission observables as a function of incident neutron energy
is crucial~\cite{nd}.
In addition, more accurate fission yields are needed to address  the
reactor anti-neutrino anomaly~\cite{RAA,RAA2}.
In major nuclear data libraries~\cite{endf,jendl,jeff,cendl},
 evaluated fission yields are only available for neutron incident energy at thermal energy, 0.5 and 14 MeV.
Both experimental measurements and reliable modelings of nuclear fission are notoriously difficult.
Nevertheless,  there was a vast amount of raw experimental fission data~\cite{exfor} but these data are generally noisy,
incomplete, and discrepant.
Therefore it is desirable to exploit the further value of these imperfect  fission data for novel nuclear applications
as well as for constraints on fission theories.

The spreading fission product yields (FPY) are the most frequently studied observables, and FPY distributions evolve in time due to cascade decays of fragments.
There are more data of cumulative fission yields (CFY) which are measured after post-fission $\beta$-decays but much fewer data of independent fission yields (IFY) which are identified promptly after fission~\cite{exfor}.
Conventionally, semi-empirical models such as Brosa~\cite{brosa} and GEF~\cite{gef} models with physics guidances are used for evaluation of
fission data. However, semi-empirical models invoke many parameters, which undermines their applicabilities for new data and uncertainty quantifications.
Advanced evaluation methods should take into account the correlations between different fission data~\cite{cgmf}.
In this context, the covariance matrix  is widely used for uncertainty assessments~\cite{endf,cov2,cov3}.

Machine learning is now revolutionizing all scientific fields and is particularly powerful for treatments of complex big data.
The belief behind data science is that data includes all correlations.
There have been increasing applications of machine learning in nuclear physics (see a review~\cite{ai}).
In particular, machine learning can be used to extract and discover knowledge from existing nuclear data, which is referred to as data mining.
For example, machine learning has been applied for inferences of nuclear structures~\cite{wn-m,utama-1,zmniu,bai,utama,xu} and nuclear reactions~\cite{ma,alovell21,Neudecker,wang}.
The application of machine learning to augment evaluations of nuclear data is very anticipated~\cite{augmented}.
Previously we have applied the Bayesian neural network (BNN) for evaluation of incomplete fission mass yields~\cite{wangza1}
and discrepant charge yields~\cite{qiao}.
The multi-layer BNN has been optimized for evaluation of fission yields with physical odd-even effects~\cite{qiao}
and non-negative constraints~\cite{wangza2}.
There was also machine learning of fission yields using the mixture density network method~\cite{alovell20}.
Besides, the Gaussian process can also solve regression problems of nuclear data evaluations but is assumed with local correlations.

In this Letter, we aim to make the best utilization of vast raw experimental fission data by data fusion~\cite{fusion}, which is a prevalent way to deal with imperfect data
and includes all  underlying
correlations. The fused data is expected to be more informative than separated data sources.  It is possible that some non-local and high-dimensional correlations could be weak but
non-negligible, in analogy to long-range and many-body interactions.
The inference would be less precise when data in some energies is sparse, however, its correlations with other data
in other energies can be beneficial to improve the inference.
It is known that Bayesian machine learning is ideal for uncertainty quantification~\cite{brand} which is key in fusion of imperfect data.
Furthermore, the fusion of complex heterogeneous data can be naturally treated by Bayesian machine learning, which exhibits the
powerful capabilities of machine learning by including correlations beyond the Gaussian process and the covariance matrix.

We employ BNN~\cite{bnn} for  machine learning of distributions of fission yields.
The posterior distribution $p(\omega |{\rm x},t)$ of BNN is based on a prior distribution $p(\omega)$ of network parameters $\omega$ and a likelihood function $p({\rm x}, t| \omega)$,
\begin{equation}
    p(\omega |{\rm x},t)=\frac{p({\rm x}, t| \omega)p(\omega)}{p({\rm x},t)}.
                                                                  \label{eqn.01}
\end{equation}
Thus the resulting inference has a distribution and naturally provides the associated uncertainty.
The data set is given as $D=\{x_i, t_i\}$, where $x_i$ is the input and $t_i$ is the output fission yields.
The details of BNN has been described in previous works~\cite{wangza1,brand}.
For the data fusion, it is crucial to taken into account the experimental uncertainties by the likelihood function,
\begin{equation}
    p({\rm x},t\mid \omega)=\exp(-\chi ^{2}/2).
                                                                  \label{eqn.02}
\end{equation}
The cost function $\chi$$^{2}$($\omega$) reads:
\begin{equation}
    \chi^{2}(\omega)=\sum^{N}_{i=1}\frac{(t_{i}-f({\rm x_{i}},\omega))^2}{\delta_{i}^2+ \sigma_{i,\rm expt}^2},                                                              \label{eqn.03}
\end{equation}
where $f({\rm x_{i}},\omega)$ denotes the network values.
Note that the  weights in the likelihood include a hyperparameter noise scale $\delta_{i}^2$ and the experimental uncertainty $\sigma_{i,\rm expt}^2$.
The hyperparameter  $\delta_{i}^2$ is uniform  for all data points and changes in the learning process until numerical convergence.
It is reasonable to see that data points with large experimental uncertainties would have small weights in the data fusion.
It is also interesting  to study how the experimental uncertainties prorogate to the prediction uncertainties.
The data fusion is in some sense similar to the conception of model averaging and model mixing~\cite{wn-m} but doesn't suffer from the bias in model selections.

\begin{figure}[htbp]
%\begin{center}
\includegraphics[width=0.49\textwidth]{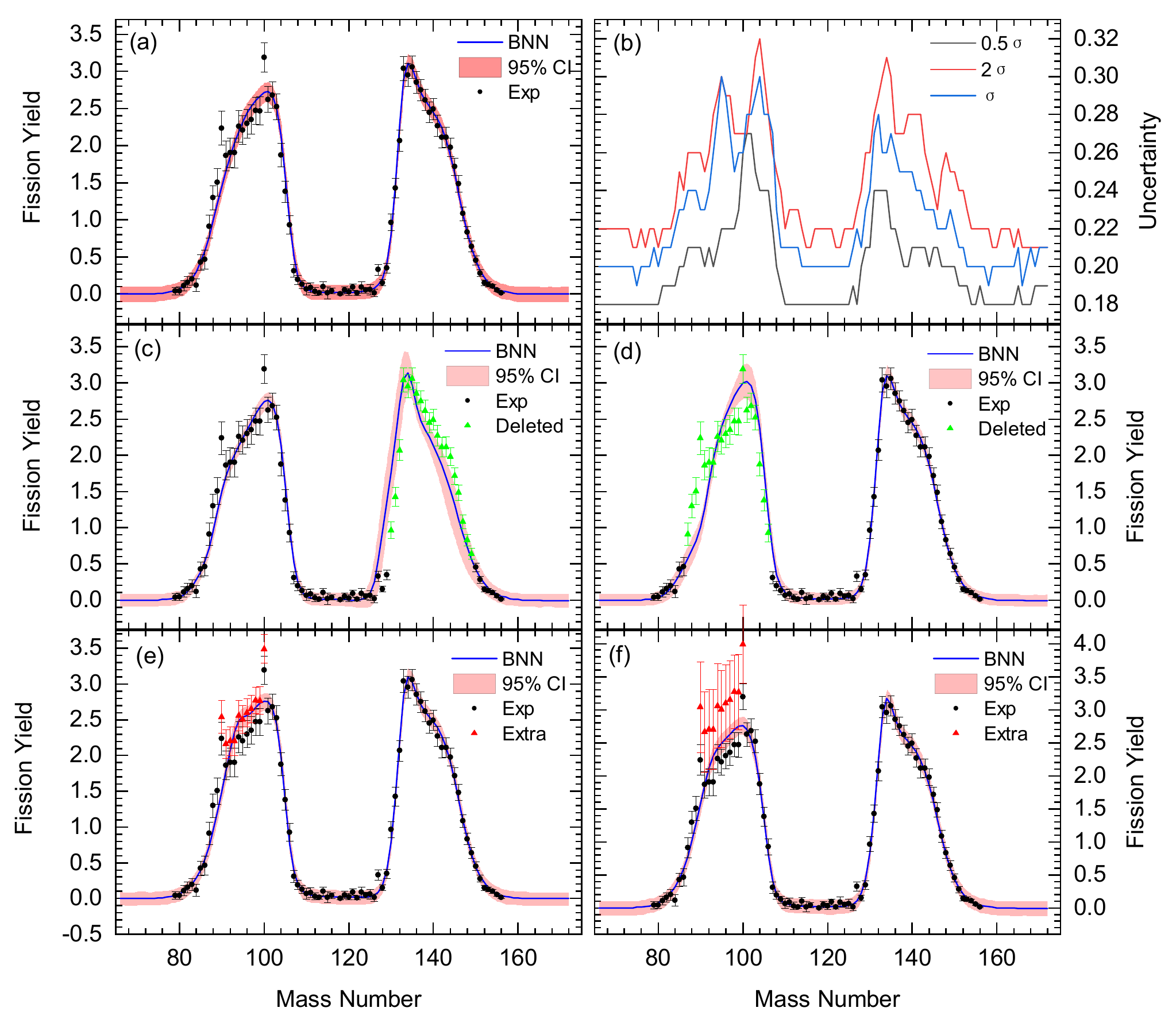}
\caption{
The BNN evaluaitons of fission  yields from neutron-induced fission of $^{238}$U, with experimental data from~\cite{Ramos}.
The shadows denote the BNN uncertainties as given by the confidential interval (CI) at 95$\%$.
(a) shows the BNN evaluation with experimental uncertainties. (b) shows the evaluated uncertainties by varying experimental uncertainties with a factor from 0.5 to 2.0.
(c) and (d) shows the evaluations when data points of the right peak or the left peak are deleted, respectively.
(f) and (g) show the evaluations with some extra noisy data (red color).
}
\label{FIG1}
%\end{center}
\end{figure}

Firstly, we demonstrate the uncertainty prorogations by varying experimental uncertainties, or by deleting or adding data points deliberately to mimic incomplete and discrepant data.
We use BNN to evaluate the fission mass distribution of $^{238}$U as an illustrative example, with experimental data from Ref.~\cite{Ramos}.
The data set also includes the evaluated independent fission yields of $^{238}$U with neutron incident energies at 0.5 and 14 MeV from JENDL~\cite{jendl}.
Note that neutron-induced fission of $^{238}$U doesn't occur at 0.5 MeV and the yield data of 0.5 MeV is a hypothetical evaluation.
Here BNN adopts a single hidden layer with 10 neurons because the tests employ a small data set of 309 data points.
Fig.\ref{FIG1}(a) displays the evaluation of fission yields of $^{238}$U  by including original experimental uncertainties.
The BNN evaluations generally agree with experimental data except that the evaluation at mass $A$=100 is lower than experiments, since
the JENDL evaluation is also lower.
Fig.\ref{FIG1}(b) shows the inferred uncertainties by varying the experimental uncertainties in Fig.\ref{FIG1}(a) with a factor from 0.5 to 2.0.
We see that uncertainties indeed increases with increasing experimental uncertainties but don't increase linearly.
It is reasonable to see that the uncertainties are generally larger around two peaks.
Fig.\ref{FIG1}(c) and Fig.\ref{FIG1}(d) show the predictions when the experimental data are deleted at the right or left peak, respectively.
We clearly see that the resulted uncertainty of the absent data is larger than that with experimental data.
Fig.\ref{FIG1}(e) and Fig.\ref{FIG1}(f) show the influences of discrepant data. In Fig.\ref{FIG1}(e), we add some extra data at the left peak that are not far from experiments with small uncertainties.
We see the the evaluation moves toward the extra data with increased uncertainties.
In Fig.\ref{FIG1}(f), we add some extra data that are far from experiments but with large uncertainties,
However, we see the evaluation doesn't move significantly towards the extra data. Therefore the uncertainty prorogation is a comprehensive effect
and would not significantly change due to a specific data.

Next we did the practical Bayesian data fusion of CFY of neutron-induced fission of $^{238}$U from different experiments with different incident energies.
The raw imperfect experimental fission yields are taken from the EXFOR library~\cite{exfor,exfor1}.
In BNN, input variables are given in terms of ($Z$, $N$, $E$), i.e., the atomic number $Z$, neutron number $N$ of fragments and
the neutron incident energy $E$. There are total 33 experiments with different energies of CFY of $^{238}$U with about 1221 scattering data points.
In addition, 2064 data points of evaluated CFY from JENDL at energies of 0.5 and 14 MeV are included as a learning constraint.
Here BNN adopts a double-layer network with 20-20 neurons.
\begin{figure}[htbp]
%\begin{center}
\includegraphics[width=0.49\textwidth]{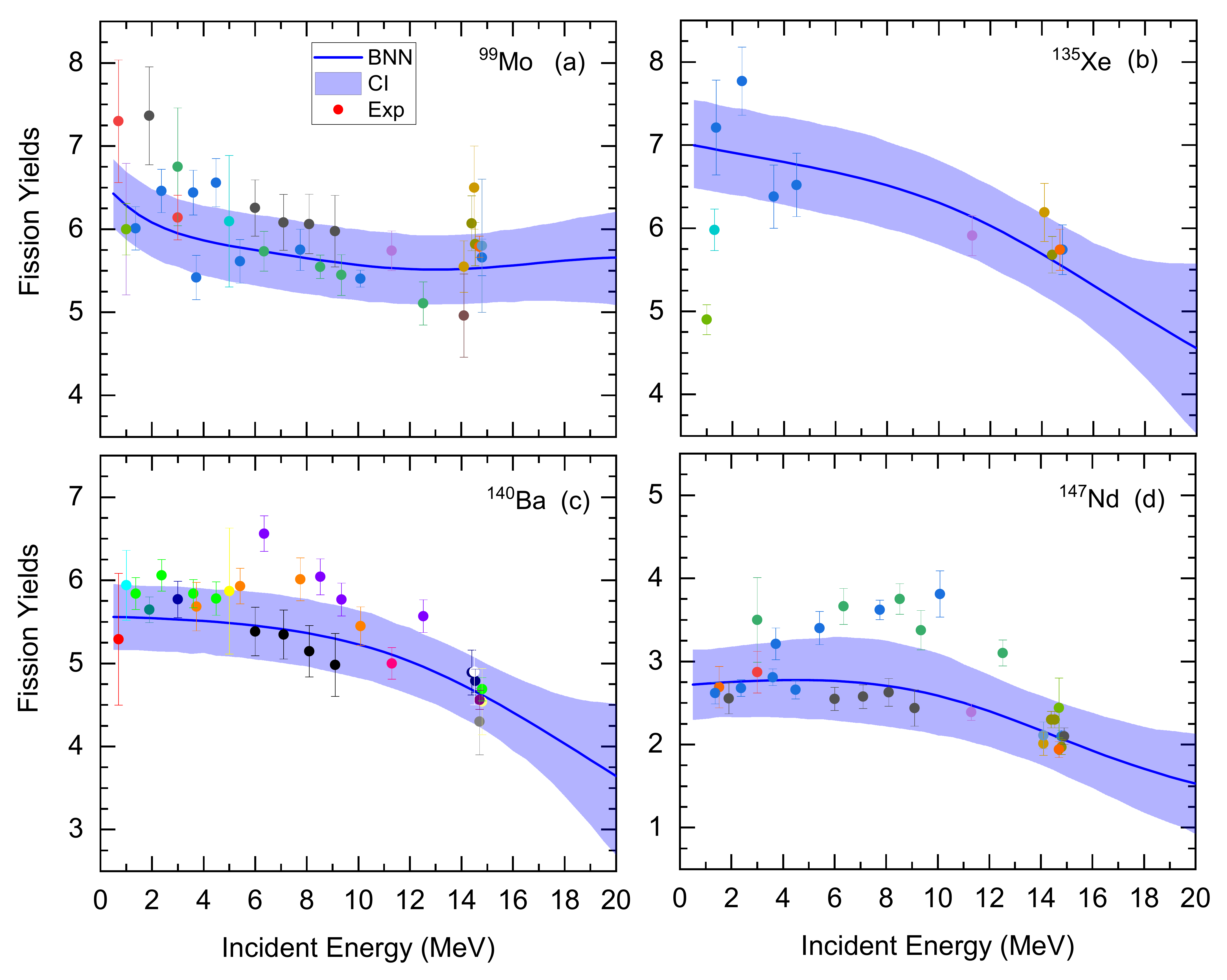}
\caption{\label{FIG2}
The BNN evaluated  yield-energy relations for $^{99}$Mo (a), $^{135}$Xe (b), $^{140}$Ba (c) and $^{147}$Nd (d) from $n$+$^{238}$U fission are shown.
The raw experimental data are taken from EXFOR~\cite{exfor}. Different experiments are denoted by different colours.
The shadows denote the corresponding BNN uncertainties given by CI at 95$\%$.
}
%\end{center}
\end{figure}

\begin{figure}[htbp]
%\begin{center}
\includegraphics[width=0.49\textwidth]{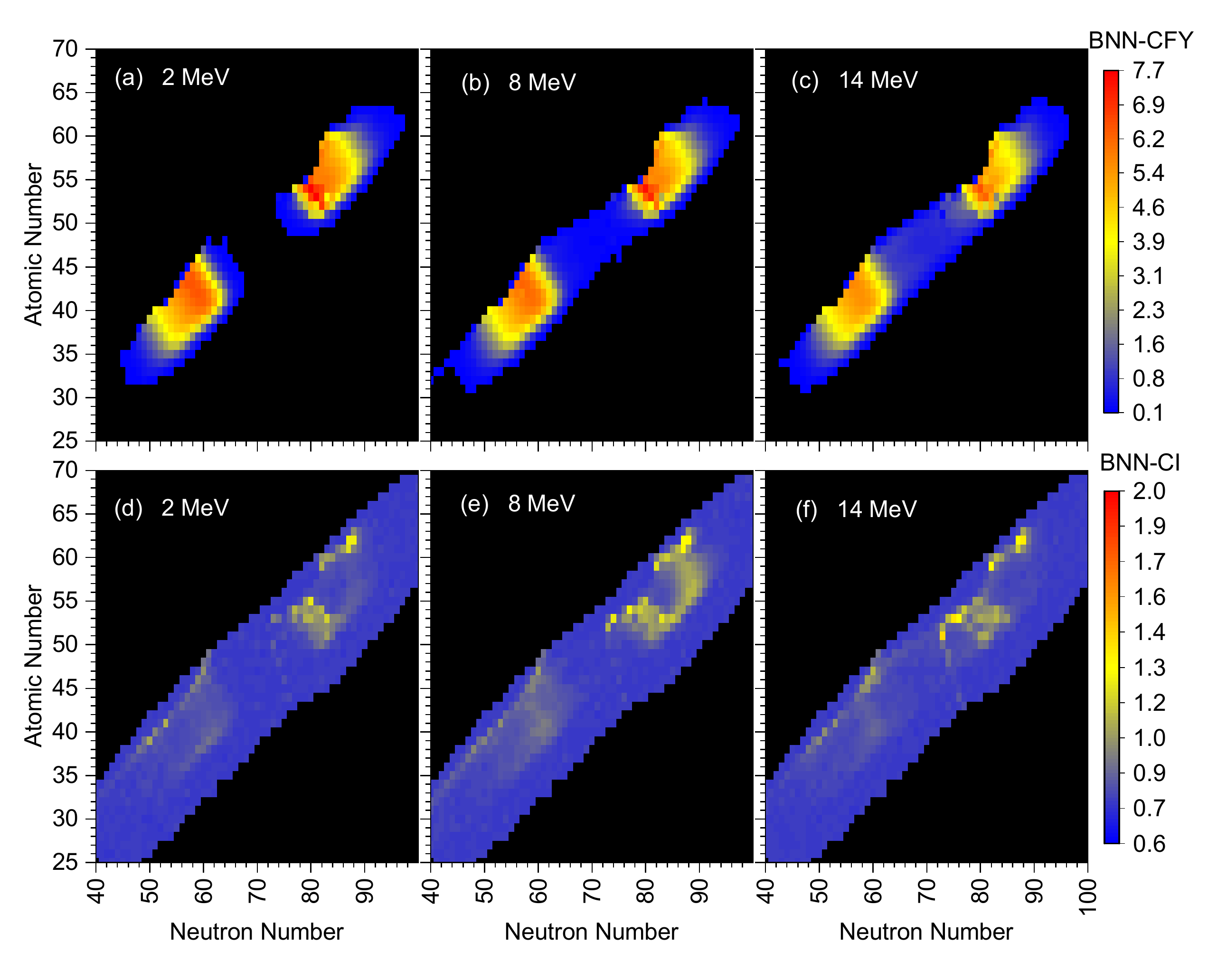}
\caption{\label{FIG3}
The two-dimensional CFY distributions of $n$+$^{238}$U fission obtained by BNN data fusion.
Panels (a), (b) and (c) shows  CFY at incident neutron energies of 2, 8 and 14 MeV respectively.
Panels (d), (e) and (f) shows the corresponding uncertainties at different energies.
}
%\end{center}
\end{figure}

The yield-energy relations of some long-lived isotopes are of particular application interests for monitoring fission environments.
For example, $^{135}$Xe has large neutron absorption cross sections and is a ``poison" for reactors~\cite{135}.
 Fig.\ref{FIG2} shows the evaluated energy dependence of fission yields of $^{99}$Mo, $^{135}$Xe, $^{140}$Ba and $^{147}$Nd.
In Fig.2, we see that the experimental data are indeed noisy, sparse and even discrepant.
We see that  data fusion can give reasonably the yield-energy relations and uncertainty quantifications.
Since JENDL evaluations are included, consequently BNN results are close to JENDL~\cite{jendl} and ENDF~\cite{endf} evaluations at 0.5 and 14 MeV  for the above 4 fragment yields.
Our key motivation is to infer the energy dependence between 0.5 and 14 MeV.
For $^{99}$Mo, CFY decreases firstly but doesn't decrease after 10 MeV.
For $^{140}$Ba, CFY decreases smoothly as energy increases.
There are more data for $^{99}$Mo and $^{140}$Ba and  they are usually adopted as standards to determine other yields.
For $^{135}$Xe, there are very few experimental data and resulting uncertainties are larger than other cases.
For  $^{147}$Nd, there are some significantly discrepant data between 5$\sim$10 MeV.
Our BNN evaluations are flat before 10 MeV and are close to the latest measurement~\cite{gooden}.
The corresponding uncertainties become larger in this energy regime.
In Fig.\ref{FIG2}, the inference uncertainties are larger than some experimental uncertainties
but they are acceptable regarding the influences of discrepant data.
The extrapolated yields at higher energies generally have increasing uncertainties.
The BNN uncertainty includes two parts: the overall regression noise and the data-dependent posterior uncertainty~\cite{bnn}, while they are not exclusive.
In Fig.\ref{FIG2}, there is an overall noise scale of $\sim$0.6 which is related to the description capability of BNN and can be reduced by
more complicated neural network or physics-informed neural network.
Besides, Fig.\ref{FIG1} and Fig.\ref{FIG2} demonstrate that the experimental uncertainty, data sparsity and data discrepancy are reflected by the data-dependent uncertainty.
Usually different experiments adopt different criterions for normalization of fission yields.
The BNN evaluations can be further improved by careful calibrations of raw experimental data.

Fig.\ref{FIG3} displays the two-dimensional CFY distributions of neutron-induced fission of $^{238}$U at energies of 2, 8 and 14 MeV from BNN data fusion.
Recent experiments have made great progress
in measurements of the isotopic mass yields using the inverse kinetic method~\cite{Ramos},
and it is still very difficult to obtain the complete distributions.
The two-dimensional CFY distribution is very different from Gaussian distributions and it is difficult to be evaluated directly
by semi-empirical models.
In principle, our results include the  yield-energy relations of all fragments.
We can see that the highest yields are around $Z$=53$\thicksim$54 and $N$=79$\thicksim$81, which decrease gradually with increasing energies.
It is known that symmetric fission would become more prominent as excitation energy increases~\cite{symf1,symf2}.
The two peaks associated with asymmetric fission modes reduce with increasing energies, while the central distributions associated with the symmetric fission
mode are increasing.
Therefore, the energy dependence of fission modes can be reasonably described by BNN data fusion.
Fig.\ref{FIG3} (d, e, f) shows the corresponding uncertainties.
There is an overall background noise scale $\sim$0.6 for all data.
Obviously the uncertainty is energy dependent. The average uncertainties of  Fig.\ref{FIG3} (d, e, f) are 0.789, 0.805 and 0.793 for 1032 points, respectively.
The larger uncertainty at 8 MeV is related to the data sparsity around this energy regime.
The uncertainties of some FPY are as large as 1.4 where CFY have large values.
The relative uncertainties around peaks are actually smaller.
The one-dimensional charge yields can also be extracted from the two-dimensional CFY distributions.
The precise and complete CFY are crucial to estimate the abnormal anti-neutrino spectrum at reactors~\cite{RAA2}
and our results will be useful for such studies.

\begin{figure}[htbp]
%\begin{center}
\includegraphics[width=0.49\textwidth]{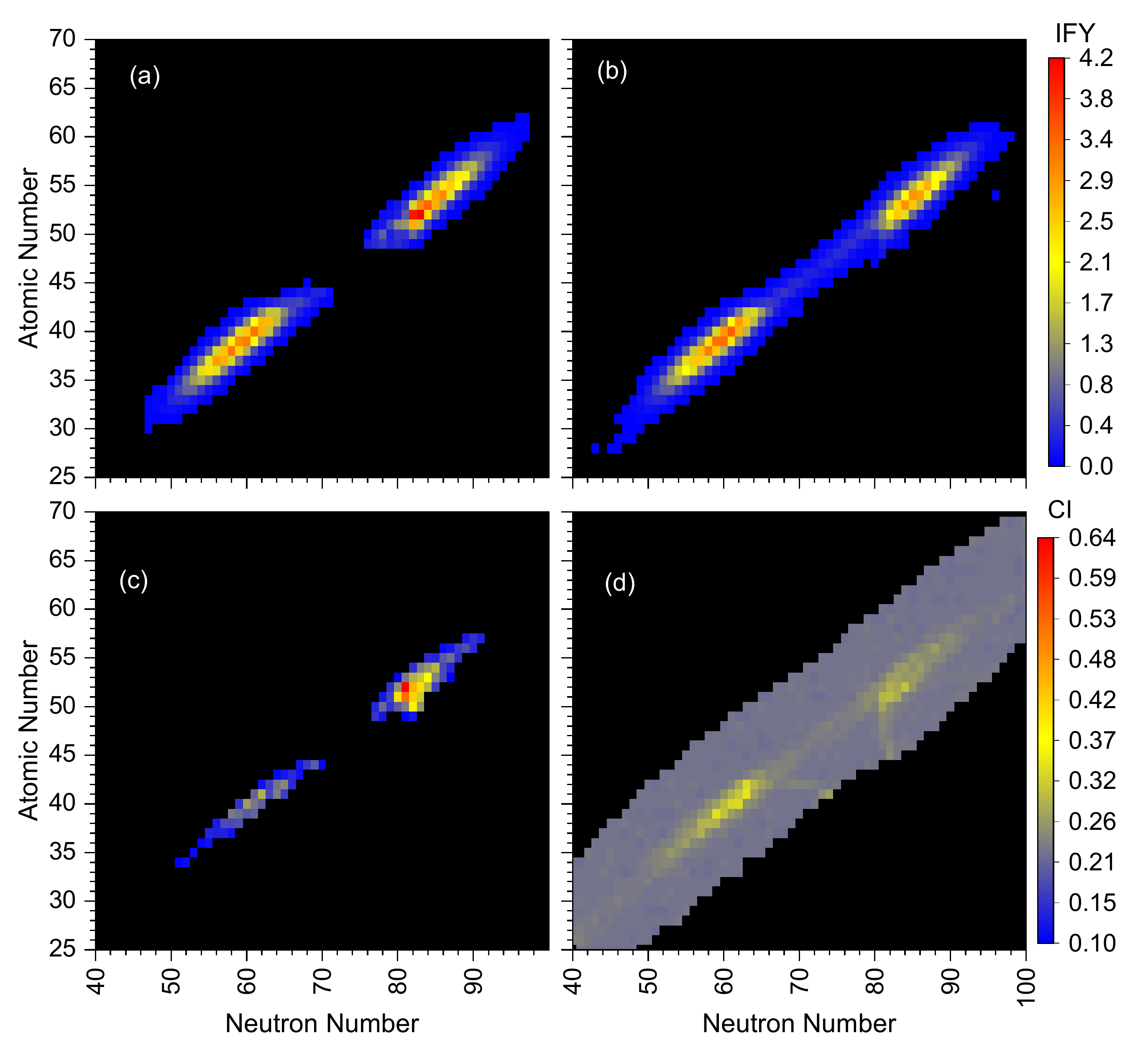}
\caption{\label{FIG4}
The two-dimensional IFY distribution of $n$+$^{238}$U fission with neutron incident energy at 8 MeV obtained by BNN data fusion.
(a) shows the inference by learning pure IFY data. (b) shows the inference by heterogeneous data fusion of CFY and IFY.
(c) and (d) shows the uncertainties corresponding to (a) and (b), respectively.
}
%\end{center}
\end{figure}

Different from CFY, there are much fewer experimental data of IFY and its evaluation is more challenging.
IFY distributions are particularly useful for constraints of fission theories.
For $^{238}$U, there are only 211 IFY  experimental data~\cite{exfor} and 2064 evaluated data points from JENDL.
For better interpolation of energy dependence of IFY, we employ the  heterogeneous data fusion
 of CFY and IFY, by utilizing the correlations between CFY and IFY.
The learning dataset is ($A_i$, $Z^{C/I}_i$, $E_i$, Y$^{C/I}_{i}$), in which
CFY and IFY share the mass number $A$ of fragments and the incident energy $E$, but their proton numbers are deliberately separated into two groups as $Z^{I}_i$ and $Z^{C}_i$.
Such a scheme is based on the fact that CFY mainly results from  $\beta$-decays of IFY but they have very different distributions.
The heterogeneous datasets are learned by the same network so that their multi-dimensional correlations are naturally included.
The evaluated IFY from JENDL at 0.5 MeV and 14 MeV are used in the learning and our main purpose is to interpolate the energy dependence at other energies.
Fig.\ref{FIG4} displays the evaluation of IFY at 8 MeV with learning of pure IFY and  heterogeneous CFY and IFY, respectively.
The homogeneous fusion of IFY adopts a 16-16 network and the heterogeneous fusion of CFY and IFY adopts a 20-20 network.
There are almost no experimental IFY data between 2 and 14 MeV.
It is known that the energy dependence of PFY is nonlinear from Fig.\ref{FIG2}.
Thus the evaluation of IFY at 8 MeV is not reliable by learning only IFY data.
We see that there are almost no FPY from the symmetric fission between two peaks in Fig.~\ref{FIG4}(a),
but there are considerable symmetric fission FPY  in Fig.~\ref{FIG4}(b).
This demonstrated that the heterogeneous data fusion of CFY and IFY can build the
energy dependence information of CFY into IFY. The resulting IFY in Fig.~\ref{FIG4}(b) are consistent with the CFY in Fig.\ref{FIG3}(b).
Fig.~\ref{FIG4}(c)(d) displays the uncertainties correspondingly.
We see that the homogeneous fusion has a much smaller overall noise but some points have large uncertainties.
The heterogeneous fusion has an overall background noise scale about 0.2 due to the influence of  CFY.
Indeed, it has been pointed out that cross-experiment correlations would result in increased final estimated uncertainties~\cite{cov3}.
On the other hand, without the background noise, the data-dependent uncertainties of  heterogeneous data fusion
are generally smaller than that of  the homogeneous fusion.

In summary, we have applied  Bayesian machine learning for the evaluation of imperfect fission yields, in which
the raw experimental data are generally noisy, incomplete, discrepant and correlated. The Bayesian data fusion can be used to
augment the inference of energy dependence of imperfect fission data by including underlying correlations, which is a crucial need for next-generation nuclear energy productions.
The two-dimensional distributions of  cumulative fission yields in terms of energy dependence are obtained.
The evolution of fission modes is revealed by the smooth transition of two-dimensional fission yields.
The yield-energy relations of key fragments are now given with uncertainty quantifications, in which
energy dependencies of uncertainties are data dependent.
We also applied the  heterogeneous data fusion to interpolate the energy dependence of independent fission yields which have
very few experimental data.
We expect this method can also be applied to the heterogeneous data fusion of multiple fission observables
and flexible physics constraints for a comprehensive understanding of nuclear fission.

%%%%%%%%%%%%%%%%%%%%%%%%%%%%%%%%%%%%%%%%%%%%%%%%%%%%%%%%%%%%%%%%%%%%%%%%%%%%%%%%
\acknowledgments
We are grateful for valuable comments by W. Nazarewicz.
This work was supported by the National Key R$\&$D Program of China (Contract No. 2018YFA0404403)
and by the National Natural Science Foundation of China under Grants
No.  11975032, 11835001, 11790325, and 11961141003.


\begin{thebibliography}{99}

\bibitem{future}
M. Bender {\it et al.}, J. Phys. G 47, 113002(2020).

\bibitem{nd}
L.A. Bernstein, D. A. Brown, A. J. Koning, B.T. Rearden, C. E. Romano, A. A. Sonzogni, A. S. Voyles, and W. Younes,
 Ann. Rev. Nucl. Part. Sci. 69, 109(2019).

%\bibitem{nd2}
%L. Bernstein, D. Brown, A. Hurst, J. Kelly, F. Kondev, E. McCutchan, C. Nesaraja, R. Slaybaugh, and A. Sonzogni, arXiv:1511.07772.

%\bibitem{sterile}
%J. Kopp, P. A. N. Machado, M. Maltoni, and T. Schwetz,
%J. High Energy Phys. 05, 050(2013).

\bibitem{RAA}
G. Mention, M. Fechner, Th. Lasserre, Th. A. Mueller, D.
Lhuillier, M. Cribier, and A. Letourneau, Phys. Rev. D 83,
073006 (2011).

\bibitem{RAA2}
A.A. Sonzogni, E.A. McCutchan, T.D. Johnson, and P. Dimitriou,
Phys. Rev. Lett. 116, 132502(2016).

\bibitem{endf}
M.B. Chadwick, et al., Nuclear Data Sheets 112,  2887 (2011).

\bibitem{jendl}
K. Shibata, et al., J. Nucl. Sci. Tech. 48, 1(2011).

\bibitem{jeff}
Joint Evaluated Fission and Fusion (JEFF)
Nuclear Data Library, https://www.oecd-nea.org/dbdata/jeff/

\bibitem{cendl}
Z.G. Ge, Z. X. Zhao, H. H. Xia, Y. X. Zhuang, T. J. Liu, J. S. Zhang and H. C. Wu,
J. Korean Phys. Soc. 59, 1052 (2011)

\bibitem{exfor}
https://www-nds.iaea.org/exfor/

\bibitem{brosa}
U. Brosa, S. Grossmann, A. M\"{u}ller, Phys. Rep. 197, 167 (1990).

\bibitem{gef}
K.H. Schmidt, B. Jurado, C. Amouroux, and C. Schmitt, Nuclear Data Sheets 131, 107(2016).


\bibitem{cgmf}
P. Talou, R. Vogt,  J. Randrup, et al.,  Eur. Phys. J. A 54, 9 (2018).

\bibitem{cov2}
K. Tsubakihara, S. Okumura, C. Ishizuka, T. Yoshida, F. Minato, and S. Chiba, J. Nucl. Sci. Tech. 58, 151(2021).

\bibitem{cov3}
P. Talou, EPJ Nuclear Sci. Technol. 4, 29 (2018).


\bibitem{ai}
P. Bedaque, A. Boehnlein, M. Cromaz, et al., Eur. Phys. J. A 57, 100 (2021)


\bibitem{wn-m}
L. Neufcourt, Y. Cao, S. A. Giuliani, W. Nazarewicz, E. Olsen, and O. B. Tarasov,
Phys. Rev. C 101, 044307 (2020).

\bibitem{utama-1}
R. Utama, J. Piekarewicz, and H. B. Prosper, Phys. Rev. C 93, 014311 (2016).

\bibitem{zmniu}
Z. M. Niu and H. Z. Liang, Phys. Lett. B 778, 48 (2018).

\bibitem{utama}
R. Utama, W.C. Chen, and J. Piekarewicz, J. Phys. G 43, 114002 (2016).

\bibitem{bai}
D. Wu, C. L. Bai, H. Sagawa, and H. Q. Zhang, Phys. Rev. C 102, 054323 (2020).

\bibitem{xu}
Y. Ma, C. Su, J. Liu, Z. Ren, C. Xu, and Y. Gao,
Phys. Rev. C 101, 014304 (2020).


\bibitem{ma}
C. W. Ma, D. Peng, H. L. Wei, Z. M. Niu, Y. T. Wang, and R. Wada, Chin. Phys.C 44, 014104 (2020).

\bibitem{alovell21}
A. E. Lovell, F. M. Nunes, M. Catacora-Rios, and G. B. King, J. Phys. G 48, 014001 (2021).

\bibitem{Neudecker}
D.Neudecker, O.Cabellos, A. R. Clark, M. J. Grosskopf, W. Haeck, M. W. Herman, J. Hutchinson, T. Kawano, A. E. Lovell, I. Stetcu, P. Talou, and S. V. Wiel,
Phys. Rev. C 104, 034611(2021).


\bibitem{wang}
Y. Wang, F. Li, Q. Li, H. Lu, K. Zhou, Phys. Lett. B 822,136669(2021).

\bibitem{augmented}
P. Vicente Valdez, L. Bernstein, M. Fratoni,
Ann. Nucl. Energy 163, 108596(2021).





\bibitem{wangza1}
Z. A. Wang, J. C. Pei, and Y. Liu, Y. Qiang, Phys. Rev. Lett. 123, 122501 (2019).

\bibitem{qiao}
C. Y. Qiao, J. C. Pei, Z. A. Wang, Y. Qiang, Y. J. Chen, N. C. Shu, and Z. G. Ge,
Phys. Rev. C 103, 034621(2021).

\bibitem{wangza2}
Z.A. Wang, J.C. Pei, arXiv:2106.11746, Phys. Rev. C (accepted).


%\bibitem{keeble}
%J. W. T. Keeble and A. Rios, Phys. Lett. B 809, 135743 (2020).





\bibitem{alovell20}
A. E. Lovell, A. T. Mohan, and P. Talou, J. Phys. G 47, 114001 (2020).

\bibitem{fusion}
T. Meng, X. Jing, Z. Yan, W. Pedrycz,
Information Fusion 57, 115(2020).

\bibitem{brand}
D.R. Phillips, R.J. Furnstahl, U. Heinz, T. Maiti, W. Nazarewicz, F.M. Nunes, M. Plumlee, M.T. Pratola, S. Pratt, F.G. Viens, and S.M. Wild,
J. Phys. G 48, 072001(2021).

\bibitem{bnn}
R. Neal,
Bayesian Learning of Neural Network,
Springer, New York (1996).

\bibitem{Ramos}
D. Ramos {\it et al.}, Phys. Rev. C 101, 034609(2020).

\bibitem{exfor1}
N.Otuka {\it et al.}, Nucl. Data Sheets 120, 272(2014).

\bibitem{135}
O. Moreira, Ann. Nucl. Energy 39, 62(2012).

\bibitem{gooden}
M.E. Gooden, et al., Nuclear Data Sheets 131,  319(2016)



\bibitem{symf1}
J. A. Sheikh, W. Nazarewicz, and J. C. Pei,
Phys. Rev. C 80, 011302(R)(2009).

\bibitem{symf2}
J. Zhao, T. Nik\v{s}i\'{c}, D. Vretenar, S.G.Zhou,
Phys. Rev. C 99, 014618 (2019).






%1234567890123456789012345678901234567890123456789012345678901234567890123456789
\end{thebibliography}
\end{document}